\documentclass[sigconf]{acmart}
\usepackage{multirow}

\usepackage{graphicx}
\usepackage{textcomp}
\usepackage{xcolor}
\usepackage{multirow} 
\usepackage{multicol} 
\usepackage{algorithm}
\usepackage{algpseudocode}

\AtBeginDocument{%
  \providecommand\BibTeX{{%
    \normalfont B\kern-0.5em{\scshape i\kern-0.25em b}\kern-0.8em\TeX}}}

\setcopyright{acmcopyright}
\copyrightyear{2018}
\acmYear{2018}
\acmDOI{XXXXXXX.XXXXXXX}

\acmConference[Conference acronym 'XX]{Make sure to enter the correct
  conference title from your rights confirmation emai}{June 03--05,
  2018}{Woodstock, NY}
\acmPrice{15.00}
\acmISBN{978-1-4503-XXXX-X/18/06}

\begin{document}
\fancyhead{}
\title{Single-shot Embedding Dimension Search in \\ Recommender System}

\author{Liang Qu}
\authornote{Both authors contributed equally to this research.}
\authornote{This work is finished when Liang Qu was an intern in WeChat, Tencent.}
\affiliation{%
  \institution{The University of Queensland}
  \city{Brisbane}
  \state{QLD}
  \country{Australia}
}
\email{l.qu1@uq.net.au}

\author{Yonghong Ye}
\authornotemark[1]
\affiliation{%
  \institution{WeChat, Tencent}
  \city{Shenzhen}
   \state{Guangdong}
  \country{China}}
\email{beardollye@tencent.com}

\author{Ningzhi Tang}
\affiliation{%
  \institution{Southern University of Science and Technology}
  \city{Shenzhen}
  \state{Guangdong}
  \postcode{518055}
  \country{China}
}

\author{Lixin Zhang}
\affiliation{%
 \institution{WeChat, Tencent}
 \city{Shenzhen}
 \state{Guangdong}
 \country{China}}
\email{lixinzhang@tencent.com}

\author{Yuhui Shi}
\authornote{Corresponding Author}
\affiliation{%
  \institution{Southern University of Science and Technology}
  \city{Shenzhen}
  \postcode{518055}
  \state{Guangdong}
  \country{China}}
\email{shiyh@sustech.edu.cn}

\author{Hongzhi Yin}
\authornotemark[3]
\affiliation{%
  \institution{The University of Queensland}
  \city{Brisbane}
  \state{QLD}
  \country{Australia}
%   \postcode{78229}
  }
\email{h.yin1@uq.edu.au}

% \affiliation{%

% \affiliation{%

\renewcommand{\shortauthors}{Liang and Yonghong, et al.}

\begin{abstract}
As a crucial component of most modern deep recommender systems, feature embedding maps high-dimensional sparse user/item features into low-dimensional dense embeddings.
However, these embeddings are usually assigned a unified dimension, which suffers from the following issues: (1) high memory usage and computation cost. (2) sub-optimal performance due to inferior dimension assignments. 
In order to alleviate the above issues, some works focus on automated embedding dimension search by formulating it as hyper-parameter optimization or embedding pruning problems. However, they either require well-designed search space for hyperparameters or need time-consuming optimization procedures.
In this paper, we propose a Single-Shot Embedding Dimension Search method, called SSEDS, which can efficiently assign dimensions for each feature field via a single-shot embedding pruning operation while maintaining the recommendation accuracy of the model. Specifically, it introduces a criterion for identifying the importance of each embedding dimension for each feature field.
As a result, SSEDS could automatically obtain mixed-dimensional embeddings by explicitly reducing redundant embedding dimensions based on the corresponding dimension importance ranking and the predefined parameter budget.  
Furthermore, the proposed SSEDS is model-agnostic, meaning that it could be integrated into different base recommendation models. 
The extensive offline experiments are conducted on two widely used public datasets for CTR (Click Through Rate) prediction task, and the results demonstrate that SSEDS can still achieve strong recommendation performance even if it has reduced 90\% parameters. 
Moreover, SSEDS has also been deployed on the WeChat Subscription platform for practical recommendation services. The 7-day online A/B test results show that SSEDS can significantly improve the performance of the online recommendation model while reducing resource consumption. 
\end{abstract}

\begin{CCSXML}
<ccs2012>
   <concept>
       <concept_id>10002951.10003317.10003347.10003350</concept_id>
       <concept_desc>Information systems~Recommender systems</concept_desc>
       <concept_significance>500</concept_significance>
       </concept>
 </ccs2012>
\end{CCSXML}

\ccsdesc[500]{Information systems~Recommender systems}

\keywords{embedding dimension search, embedding pruning, recommender system, sparse learning}

\maketitle

\section{Introduction}
Recommender systems have been widely deployed to various scenarios such as advertisement \cite{zhou2018deep}, online shopping \cite{cheng2016wide}, news apps \cite{zheng2018drn}, and many others \cite{nguyen2017argument,wang2020next,chen2021learning,chen2020multi,zhang2018discrete,yin2016spatio}. 
The typical inputs of recommender systems are a large number of categorical (e.g., gender) or numerical (e.g., age) features associated with users and items. For example, in WeChat and Youtube platforms, more than a billion unique user ID features are encoded as high-dimensional sparse one-hot feature vectors. To well extract users' preferences for personalized recommendations, most state-of-the-art 
recommendation methods, such as deep neural network (DNN) based methods \cite{cheng2016wide, covington2016deep}, factorization machine (FM) based methods \cite{rendle2010factorization, guo2017deepfm, lian2018xdeepfm, xiao2017attention}, map these high-dimensional sparse feature vectors into low-dimensional dense embeddings. Then these embeddings are utilized for further feature operations (e.g., feature interactions) to make final predictions. However, most of these methods set a fixed embedding dimension for all features, which could suffer from the following issues: (1) The embeddings could contain tens of billions of parameters resulting in high memory usage and computation cost. (2) Over-parameterizing the low-frequency features might induce overfitting and even unexpected noise. On the other hand, high-frequency features need more parameters to convey fruitful information. 

Nevertheless, manually setting appropriate embedding dimensions for different features is intractable due to the vast amount of candidate solutions. Thus, it is natural to think about how to assign embedding dimensions to different features in an automated manner, which is termed as embedding dimensions search (EDS) problem in this paper. 

The early work \cite{ginart2021mixed} tries to address EDS by introducing a human-designed rule which assigns embedding dimensions according to the popularity of features.
Recently, inspired by the success of neural architecture search (NAS) \cite{zoph2016neural},
some works employ NAS-based methods to handle EDS by formulating it as a hyper-parameter optimization (HPO) problem \cite{zheng2022automl}. For example, NIS \cite{joglekar2020neural} and ESAPN \cite{liu2020automated} search embedding dimensions from a set of predefined candidate dimensions by policy networks.
DNIS \cite{cheng2020differentiable} and AutoEmb \cite{zhao2020autoemb} utilize the differential architecture search (DARTS) \cite{liu2018darts} method to enforce the search efficiency. However, this kind of method generally requires a well-designed search space for candidate embedding dimensions and expensive optimization processes to train the candidates.
As an alternative solution to the EDS problem, some works \cite{liu2021learnable,yan2021learning,deng2021deeplight} treat EDS as an embedding pruning problem, eliminating requirements for predefining search space. Instead, they obtain the mixed-dimensional embeddings by identifying and removing the redundant embedding dimensions via additional mask layers with learnable threshold parameters. 
However, the embedding pruning based EDS methods need alternatively optimize threshold parameters and parameters of the model itself, which is time-consuming, thereby undermining their utility in practical recommendation services.

To alleviate the limitations mentioned above of embedding pruning based EDS methods, this paper needs to address the following challenges: 
a) \textbf{How to identify embedding dimensions for various feature fields?}
The EDS problem could be transformed into identifying the importance of each dimension of embeddings. In this way, we could prune those relatively unimportant dimensions of embeddings such that the mixed dimension embeddings are automatically obtained.
To this end, inspired by the weights pruning method \cite{lee2018snip} in DNN, we address this challenge in a data-driven manner, which introduces a criterion that could identify the importance (called saliency score) of each embedding dimension for each feature field based on its influence on the loss function. Specifically, we mask each embedding dimension of each feature field while keeping the others unchanged. Then we could compute the saliency scores of each dimension by measuring its influence on the loss function. 
b) \textbf{How to search embedding dimensions in an efficient way?}
Based on the above idea, once we obtained the saliency scores, we could rank the embedding dimensions over all feature fields in descending order and retain a limited portion based on the given parameter budget.
However, it is prohibitively expensive to identify the importance of each dimension one by one. Thus, we utilize an approximation operation \cite{koh2017understanding} to efficiently measure the importance of all dimensions only in one forward-backward pass, namely single-shot embedding pruning. Hence, one significant advantage of SSEDS is its efficiency, which makes it suitable for practical industry recommender systems requiring frequently (e.g., per hour) updating models due to the real-time changes in feature distribution.       
c) \textbf{How to integrate the proposed SSEDS into traditional recommendation models?} The traditional recommendation models, especially FM based methods \cite{rendle2010factorization, guo2017deepfm, lian2018xdeepfm, xiao2017attention}, utilize explicit feature interaction operations (e.g., the dot product) to capture cross feature relations, requiring all embeddings to have the same dimension. To make the proposed SSEDS be seamlessly integrated into various traditional models, we propose first to pretrain the traditional model in a standard way, and then obtain a slim model with mixed-dimensional embeddings via the proposed single-shot embedding pruning. Finally, we initialize and retrain the slim model using the pretrained and mixed-dimensional embeddings, and introduce additional transform matrices for each feature field to align dimensions for further feature interaction operations.

In summary, the main contributions of this paper are as below:
\begin{itemize}
    \item This paper proposes an effective and efficient single-shot embedding dimension search method, called SSEDS, which introduces a criterion that could identify the importance of each embedding dimension of each field only in one forward-backward pass. In this way, the mixed-dimensional embeddings are efficiently obtained. 
    \item The proposed SSEDS is model-agnostic. It proposes to utilize linear transform matrices to align the various dimensions for different feature fields such that SSEDS could be seamlessly integrated into various base recommendation models. On the other hand, it can flexibly control the number of pruned parameters to satisfy different requirements on the parameter budget.
    \item The extensive offline experiments are conducted on two public datasets for the CTR prediction task, and the experimental results demonstrate that SSEDS can still achieve strong recommendation performance even if it has reduced 90\% parameters. Furthermore, SSEDS has also been deployed on the WeChat Subscription platform for practical recommendation service, and the 7-day online A/B test results show that SSEDS can significantly improve the performance of the online recommendation model while reducing resource consumption.
\end{itemize}

The rest of this paper is organized as follows. Section 2 reviews the main related work. Section 3 formulates the problem and details the proposed SSEDS. Section 4 introduces experimental settings and discusses experimental results, followed by a conclusion in Section 5. 

% to do
% to summarize three challenges about SSEDS
% to summarize three contributions of this paper
% to introduce the difference between traditional AutoML and AutoML for recommender system

\section{Related Work}
This section introduces the main related works to our study, including feature-based recommendation models, embedding dimension search methods, and network pruning methods.

\subsection{Feature-based Recommender System}
% Most modern recommendation models take user/item raw high-dimensional sparse features as inputs,
% (these raw features are either one-hot or multi-hot vectors of high dimension)
% and map them into a low-dimensional dense embedding space to capture users' preference for personalized recommendations. 
% the objective is to predict whether the user will interact with the target item.
% The research methods are abundant, from shallow model linear regression (LR) to deep models such as Wide \& Deep \cite{cheng2016wide}, DeepFM \cite{guo2017deepfm} and XDeepFM \cite{lian2018xdeepfm}. Besides, attention mechanism has been applied to explore more semantic features, e.g., AFN \cite{covington2016deep}, AutoInt \cite{song2019autoint}, InterHAt \cite{li2020interpretable} and so on. 

The feature-based recommender system takes the high-dimensional and sparse features from user and item as inputs and maps them into a low-dimensional dense embedding space to better capture users' preference for personalized recommendations. To generate effective representations, deep models are widely used and provide state-of-the-art results \cite{zhang2019deep}. The related works are flourishing. The linear regression (LR) model gets extensive applications in the early stage, which directly maps the raw features to continuous predictions via a single fully-connected layer. Then Wide$\&$Deep \cite{cheng2016wide} introduces an extra MLP branch for supplementing the high-level representation. Furthermore, DeepFM \cite{guo2017deepfm}, and XDeepFM \cite{lian2018xdeepfm} turn eyes on modeling the concurrence of different features and propose factorization machine (FM). Recently, more complex deep neural layers are adopted in recommendation system, e.g., attention-based models such as AFN \cite{covington2016deep}, AutoInt \cite{song2019autoint} and InterHAt \cite{li2020interpretable}.     
However, these methods assign a fixed embedding dimension for all features regardless of their heterogeneity, which could downgrade the model performance and consume huge amount of storage and computing resources.

% \subsection{AutoML for Recommendation Model}
% The current research work on AutoML for recommendation model could be categorized into two classes, namely % automated embedding dimension search and automated feature interaction search.

\subsection{Embedding Dimension Search Methods  }
Studies on the EDS problem could be categorized into heuristic methods, hyper-parameter optimization (HPO) methods, and embedding pruning methods.
The heuristic method such as MDE \cite{ginart2021mixed} allocates the embedding dimensions based on the popularity of features. However, using such simple rules to determine the embedding dimension suffers a loss of generality for various recommendation tasks.

Recently, inspired by the success of neural architecture search (NAS) \cite{zoph2016neural,elsken2019neural}, some works model the EDS problem as HPO problems which automatically search embedding dimensions from a predefined candidate dimension set. For example, NIS \cite{joglekar2020neural} is the first work to formulate the EDS as an HPO problem, which optimizes the assignment for embedding dimensions by constantly improving the policy network with high cost on training time. Differently, some research work \cite{cheng2020differentiable, zhao2020memory, liu2020automated, zhao2020autoemb} propose to use the differentiable architecture search (DARTS) \cite{liu2018darts} to enforce the search efficiency. For example, DNIS \cite{cheng2020differentiable} adopts a soft layer to control the significance of each embedding dimension and prunes the unimportant components after training. AutoDim \cite{zhao2020memory} utilizes a soft and continuous manner to calculate the weights over various dimensions for feature fields, then the embedding architecture is derived from the maximum weight. Besides, ESAPN \cite{liu2020automated} and AutoEmb \cite{zhao2020autoemb} dynamically update the embedding structure for users and items regarding the on-time frequency as an important reference. However, HPO based EDS methods generally require well-designed search space for candidate embeddings and need iterative optimization procedures throughout training, thereby undermining their utility in practical recommendation services requiring high efficiency.

\subsection{Network Pruning}
Differently, instead of requiring predefined search space, embedding pruning based EDS methods selectively remove redundant embedding dimensions by introducing additional mask layers with learnable threshold parameters. For example, PEP \cite{liu2021learnable} designs an adaptive threshold to filter out the redundant embedding dimensions with low magnitude, and ATML \cite{yan2021learning} calibrates the breaking point to identify the promising elements. Deeplight \cite{deng2021deeplight} proposes to prune both parameters in the embedding layer and DNN layer to solve the high-latency issues in CTR prediction. However, these methods either fail to reach high efficiency \cite{liu2021learnable,yan2021learning,deng2021deeplight} due to the iterative optimization procedures or cannot strictly constrain the sparsity level \cite{liu2021learnable,yan2021learning} required by various infrastructures.

\begin{figure*}[htbp]
\centering
\setlength{\abovecaptionskip}{0.2cm}
\includegraphics[width=1.0\textwidth]{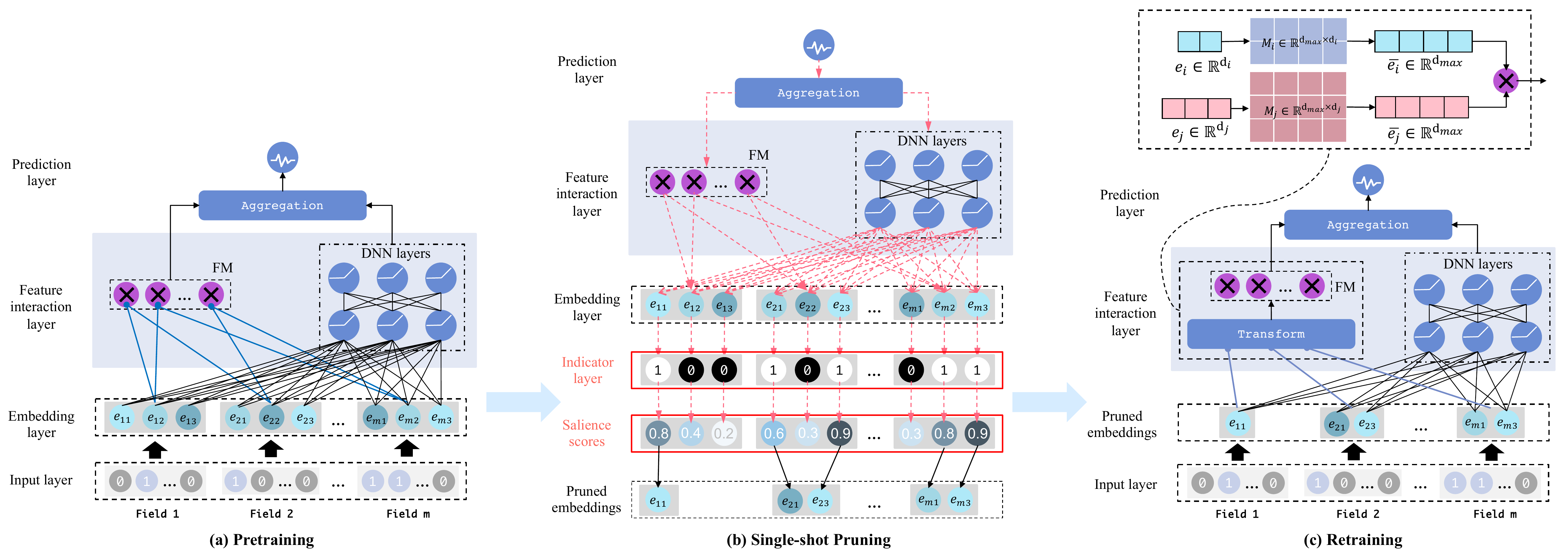}
\caption{The overview of SSEDS. a) The pretraining process trains traditional base models with uniform-dimensional embeddings. b) The single-shot pruning process identifies the salience scores of each dimension of each field and outputs mixed-dimensional embeddings. c) The retraining process aligns the dimension for different fields and retrains the new slim model in a standard way.} %最终文档中希望显示的图片标题
\label{overview} %用于文内引用的标签
\vspace{-0.3cm}
\end{figure*}

\section{Proposed Method}
This section will first give a problem formulation of the embedding pruning based EDS under the given parameter budget $\kappa$ for the feature-based recommender system\footnote{We focus on the CTR prediction task in this paper} and then elaborate on the proposed SSEDS. 

\subsection{Problem Formulation}
The typical training data $\mathcal{D}$ of the feature-based recommender system is commonly constructed from the users' click records, and each record $(\mathbf{x},y) \in \mathcal{D}$ is denoted as:
\begin{equation}
    (\mathbf{x}, y) = (\{ \mathbf{x}_{1}, \mathbf{x}_{2},...,\mathbf{x}_{m}\}, y)
\end{equation}
where $\mathbf{x} = \{\mathbf{x}_{1}, \mathbf{x}_{2},...,\mathbf{x}_{m}\}$ is the raw feature vector that concatenates $m$ feature fields associated with users and items, and $y$ is the binary label (e.g., 1 for click and 0 for not click) describing the user's preference to the given item. For each feature field (e.g., occupation), it contains a certain number of unique features (e.g., teacher, doctor and so on), thus the feature vector $\mathbf{x}_{i} \in \mathbf{x}$ of the $i$-th field is usually encoded as the high-dimensional sparse one-hot or multi-hot vector. To well extract the user's preference, most modern feature-based recommender systems map $\mathbf{x}_{i} \in \mathbb{R}^{n_{i}}$ ($n_{i}$ is the number of unique features in field $i$) into a low-dimensional dense embedding $\mathbf{e}_{i}$ as follows:
\begin{equation}
    \mathbf{e}_{i} = \mathbf{V}_{i}\mathbf{x}_{i} \label{eq:emb_lookup}
\end{equation}
where $\mathbf{V}_{i} \in \mathbb{R}^{d \times n_{i}}$ is the embedding table of the $i$-th feature field, and $d$ is the embedding dimension shared by all fields. The whole embedding table $\mathbf{V} = \{\mathbf{V}_{1},\mathbf{V}_{2},...,\mathbf{V}_{m}\}$ could be obtained by concatenating all the embedding table from each field.

Thus, the goal of a feature-based recommender system $f$ is to predict the probability $p$ about whether the user would click on the item as follows:
\begin{equation}
    p = f(\mathbf{x}|\mathbf{V}, \Theta) \label{eq:forward_pass}
\end{equation}
where $\Theta$ represents other parameters of the model (e.g., parameters in deep layers of the Wide$\&$Deep \cite{cheng2016wide} and DeepFM \cite{guo2017deepfm}).

% where $\mathbf{x}^{[i]}=\{\mathbf{x}^{[i]}_j\in [0,1]^{N_j^x}, j=1,...,m\}$ are sparse one-hot or multi-hot vectors containing $m$ fields of user/item features, and $y^{[i]}$ is the ground-truth label indicating whether the user prefer the given item. 
% A deep recommendation model, paramterized by $\Theta$, takes $\mathbf{x}^{[i]}$ as inputs and tries to generates the prediction proximal to $y^{[i]}$. A crucial component of $\Theta$ is the shard embedding table $\mathbf{E}=\{\mathbf{E}_{j} \in \mathbb{R}^{N_j^x\times d}, j=1,...,m\}$ with size $d$, which maps each input $\mathbf{x}_j^{[i]}$ to the low-dimensional dense vector $\mathbf{e}_j^{[i]}$ via embedding lookup operation:  $\mathbf{e}_j^{[i]}=\mathbf{E}_{j}^{\text{T}}\mathbf{x}_j^{[i]}$, where $\mathbf{e}_j^{[i]}\in \mathbb{R}^d$. The whole embeddings $\mathbf{e}^{[i]}=||_{i=1,...,m}\mathbf{e}_j^{[i]}$ with size $m\times d$, where $||$ represents the concatenation operation along the last dimension. 
Finally, in order to obtain mixed-dimensional embeddings, the embedding pruning based EDS methods aim to identify and remove a certain number of embedding dimensions from the whole embedding table $\mathbf{V}$ based on the parameter budget $\kappa \in (0,1]$, while minimizing the loss function as follows:
\begin{equation}
\mathbf{V}^{*},\ \mathbf{\Theta}^{*} = \mathop{\arg \min}\limits_{\mathbf{V}, \mathbf{\Theta}} \ \mathcal{L}(\mathbf{V},\Theta; \mathcal{D}), \ s.t.\  ||\mathbf{V}^{*}||_{0} < \kappa ||\mathbf{V}||_{0} \label{eq:problem}
\end{equation}
where $\mathcal{L}$ is the loss function (e.g., cross-entropy) , and $||\cdot||_0$ represents the $L_{0}$ norm, i.e., the desired non-zero parameters in final optimized pruned embedding table $\mathbf{V}^{*}$.
% = \{\mathbf{V}^{*}_{1},\mathbf{V}^{*}_{2},...,\mathbf{V}^{*}_{m}\}$,$\mathbf{V}_{i}^{*} \in \mathbb{R}^{d_{i} \times n_{i}}$, where $d_{i}$ is the learned embedding dimension for $i$-th field.

It is worth noting that the pruning under the recommender system scenario is different from the traditional network pruning methods introduced in Section 2.3. The main difference is that our method focuses on pruning embedding table $\mathbf{V}$ as it dominates the vast majority of parameters of the model, instead of the model parameters $\Theta$ focused by traditional network pruning methods. Nevertheless, it is easy to extend our method to 
pruning both embedding table and model parameters.
% Equation (\ref{eq:problem}) covers the situation that the embedding size for each concrete feature instance is specified. However, we refrain that features under one field share the same embedding size, which greatly reduces the solution space and thus benefits for efficient optimization. What's more, the sparsity for each embedding $\mathbf{e}^{[i]}$ in $\mathcal{D}$ stays the same, thus we can convert the sparsity constraint on $\mathbf{E}$ to any $\mathbf{e}^{[i]}$. In the following description, we will eliminate the upper index $^{[i]}$ and focus on an arbitrary instance $(\mathbf{x}, y)$.

% In the following subsections, we will dismiss the upper instance index $^{[i]}$ and focus our descriptions on a single instance.

% The number of  Notice that features under one field share the same embedding size, thus the pruning is executed on field level. Specifically, we introduce a mask layer with learnable binary parameters $\boldsymbol{\alpha}_j \in [0,1]^d, j=1,...,m$ to prune $\bold{e}_{j}^{[i]}$ via broadcast element-wise multiplication, resulting in the rectified model parameters $\Theta(\boldsymbol{\alpha})$. Overall, the objective function is shown as below:

%\begin{equation}
%\begin{split}
%    \mathop{\min}\limits_{\Theta, \boldsymbol{\alpha}}&\  %\mathcal{L}(\Theta(\boldsymbol{\alpha});\mathcal{D}) \\
%    s.t.&\  \Vert \boldsymbol{\alpha} \Vert_0 < \kappa
%\end{split}
%\end{equation}
% \Theta^{*}(\Omega) = \text{argmin}_{\Theta}\mathcal{L}((\Theta(\Omega), \Omega); \mathcal{D}_{train})

\subsection{SSEDS}
Figure \ref{overview} displays the overview of the proposed SSEDS that contains three stages: pretraining, single-shot pruning, and retraining. Concretely, in the pretraining stage, we pretrain those traditional recommendation models with uniform-dimensional and over-parameterized embeddings in a standard manner to make these embeddings expressive. In the single-shot pruning stage, we first compute the saliency scores (more on this later) for each embedding dimension for each field based on its influence on the loss function, and then the saliency scores are ranked in descending order. Thus we can sequentially remove dimensions with low saliency scores until the parameter budget is reached. In this way, the mixed-dimensional embeddings are automatically obtained.
In the retraining stage, since feature interaction operations (e.g., the dot product) require embeddings to have the same dimension, we propose to utilize additional transform matrices to align dimensions for all fields, as shown in Figure \ref{overview}(c). In this way, the obtained mixed-dimensional embeddings could be seamlessly integrated into architectures of traditional base models, which can be retrained in a standard way. The detailed procedures of SSEDS are summarized in Algorithm \ref{alg:SSEDS}.

% with two addictive mask layers plugged into some general FM models like FM and DeepFM (Figure \ref{overview}(a)), namely the "embeddings mask" layer (boxed by a red rectangle at the bottom) and the "interactions mask" layer (boxed by a red rectangle at the top). Specifically, a trainable embeddings mask layer for automatically setting embedding size is built on top of the classical embedding layer, where only the important components marked by the corresponding mask elements with value 1 (white circles) will be retained, resulting in "pruned embeddings" with various size. Furthermore, feature interactions are only operated among the pruned embeddings with the same size, some of which are still redundant. To promote the quality of feature interactions, another trainable interaction mask layer is introduced to discard the abundant ones. Besides, the pruned embeddings are fed to other deep layers like multi-layer perceptrons (MLP) \cite{guo2017deepfm}, attention mechanism \cite{li2020interpretable} and graph neural network (GNN) to gain deep representations. Finally, all these feature representations are concatenated to make the prediction. In the followings, we will describe these stages mathematically.

% overview..., as shown in Figure 2.
% components:
% 1. embedding mask layer
% 2. interaction layer

% (the multi-hot vector will be mapped to multiple embeddings with size $d$ at first, then these embeddings are averaged to a single one) 

\subsubsection{\textbf{Pretraining}}
It is worth noting that SSEDS is model-agnostic, which can be employed to various recommendation models like FM \cite{rendle2010factorization}, Wide$\&$Deep \cite{cheng2016wide} and DeepFM \cite{guo2017deepfm}. Therefore, we will not roll out the details of these uniform-dimension based recommendation methods.
Instead, we will introduce the general training steps of these models. Specifically, we need to pretrain a complete recommendation model in a number of iterations before pruning, which aims at making the elements in the embedding table $\mathbf{V}$ expressive. The forward pass procedures of the base recommendation model are abstracted as three layers including the embedding layer, the feature interaction layer, and the prediction layer.  

For each training instance $(\mathbf{x},y)$, the embedding layer takes the raw features $\mathbf{x} = \{\mathbf{x}_{1}, \mathbf{x}_{2},...,\mathbf{x}_{m}\}$ as inputs, and maps it into the embeddings $\mathbf{e} = \{\mathbf{e}_{1},\mathbf{e}_{2},...,\mathbf{e}_{m}\} $ in a field-wise manner via the Formula (\ref{eq:emb_lookup}): $\mathbf{e}_{i} = \mathbf{V}_{i}\mathbf{x}_{i}$.

% Concretely, the model takes $\mathbf{x}_j\in [0,1]^n_j$ as inputs, which is mapped to dense embeddings $\mathbf{e}_j \in \mathbb{R}^d$ via embedding lookup operation as Equation (\ref{eq:emb_lookup}). The whole embedding $\mathbf{e}$ is obtained by concatenating (denoted by symbol "||") all $\mathbf{e}_j, j=1,...,m$:
% \begin{equation}
%   \mathbf{e}=||_{i=1,...,m}\mathbf{e}_j
% \end{equation}

Then the feature interaction layer is utilized to capture the implicit interacted relations among these embeddings $\mathbf{V}$ by the feature interaction operations $g(\cdot)$ (e.g., the FM \cite{rendle2010factorization,guo2017deepfm} and the DNN \cite{guo2017deepfm, cheng2016wide}) as follows: 
% transform the embedding $\mathbf{e}$ to the deep representation $\mathbf{z}$, which is denoted as function $r$:
\begin{equation}
    \mathbf{z} = g(\mathbf{V}, \Theta)
\end{equation}

where $\mathbf{z}$ is the logit which will be fed into the final prediction layer (e.g., a sigmoid function) to obtain the prediction probability $p$ as follows:
% , the probability $p$ is obtained by feeding $\mathbf{z}$ to the prediction layer $h$ (e.g., a single fully-connected layer) as follows:
\begin{equation}
    p = sigmoid(\mathbf{z})
\end{equation}

Recall that the learnable parameters in the embedding layer and the feature interaction layer are $\mathbf{V}$ and $\Theta$, respectively. The loss function $\mathcal{L}(V,\Theta;\mathcal{D})$ (e.g., the cross-entropy loss) is designed to measure the discrepancy between the model prediction probability $p$ and the ground-truth label $y$, which is formulated as follows:
\begin{equation}
    \mathcal{L}(V, \Theta; \mathcal{D}) = -\sum_{\{\mathbf{x}, y\}\in \mathcal{D}} \{y\cdot \log(p)+(1-y)\cdot \log(1-p)\}
\end{equation}

By minimizing the loss $\mathcal{L}(V, \Theta; \mathcal{D})$, we can obtain the optimized embedding table $\hat{\mathbf{V}}$ with the uniform embedding dimension $d$ for all feature fields, and other optimized parameters $\hat{\mathbf{\Theta}}$. 
% Next we should prune $V^{*}$ to assign various embedding size for different features.
% In the following sections, we will decompose $\Theta$ to two parts: $\Theta^{\prime}$ and $\mathbf{E}$ for better statement, i.e., $\Theta^{\prime}$ denotes for all trainable parameters except for $\mathbf{E}$ in $\Theta$. Meanwhile, the optimal $\Theta^{\prime}$ and $\mathbf{E}$ after pretraining are defined $\Theta^{\prime}^*$ and $\mathbf{E}^{*}$, repspectively.

% $\hat{y}=f(\mathbf{x}; \Theta^{\prime}, \mathbf{E})$

\subsubsection{\textbf{Single-shot Pruning}}
Recall that the EDS problem could be transformed into the embedding pruning problem, i.e., identifying and removing those relatively unimportant dimensions of embeddings such that mixed-dimensional embeddings are automatically obtained. Hence, the critical problem is identifying the importance (a.k.a. saliency scores) for each dimension. Existing methods \cite{liu2021learnable,yan2021learning} introduce the additional mask layer with learnable threshold parameters to learn the importance of embedding dimensions. However, they need alternatively optimize threshold parameters and parameters of the model itself, resulting in expensive training processes.
% which is often realized by measuring the saliency scores, i.e., the impact on the performance of the model once the embedding dimension is dropped. 
% The definition of saliency scores varies, for example, some work \cite{han2015learning, narang2017exploring} directly use the magnitude of the parameters, while the work \cite{wang2020picking} uses the magnitude of the Hessian gradients with respect to the parameters which reflect the impact on the loss function. However, these kinds of saliency scores are easily affected by the normalization layers. Furthermore, they are required to be alternatively optimized with other model parameters. 
Inspired by SNIP \cite{lee2018snip}, we introduce a saliency criterion to identify the importance of each embedding dimension of $\hat{\mathbf{V}}_{ij}$ for each field $i \in \{1,\cdots,m\}$ and each dimension $j \in \{1,\cdots,d\}$ independently. 
In particular, we introduce an auxiliary indicator layer with binary parameters $\boldsymbol{{\alpha}} = \{\boldsymbol{{\alpha}}_{1},\boldsymbol{{\alpha}}_{2},...,\boldsymbol{{\alpha}}_{m} \}, {\boldsymbol{\alpha}}_i\in [0, 1]^{d \times n_{i}}$. 
% for the embedding $\hat{\mathbf{e}}$ whose each component $\hat{\mathbf{e}}_{i} = \hat{\mathbf{V}}_{i}\mathbf{x}_{i}$ . Recall that we prune the embeddings in the field-level, i.e., the dimension of embeddings in the same feature field are pruned to be the same. 
% Then the sparse embedding $\tilde{\mathbf{e}}$ is obtained by element-wise production $\tilde{\mathbf{e}} = \hat{\mathbf{e}} \odot \boldsymbol{\alpha}$. 
Then, for the desired parameter budget $\kappa$, 
% the sparsity level of the embedding ${\tilde{\mathbf{e}}}$ is determined by the sparsity level of the binary matrix $\boldsymbol{\alpha}$, which allows us to
we can reformulate the objective function listed in Equation (\ref{eq:problem}) as follows:
\begin{equation}
    \begin{split}
        \mathop{\min}\limits_{\hat{\mathbf{V}},\hat{\Theta}},&\ \mathcal{L}(\hat{\mathbf{V}}\odot \boldsymbol{\alpha},\hat{\Theta}; \mathcal{D}) \\
        s.t.&\ \boldsymbol{\alpha} \in \{0, 1\}^{d \times \sum_{i}^{m}n_{i}},\ \Vert \boldsymbol{\alpha} \Vert_0 < \kappa ||\mathbf{V}||_{0} \label{eq:pruned_loss}
    \end{split}
\end{equation}

% Obviously the mask $\boldsymbol{\alpha}$ paves the way for studying the impact on the loss function if we drop the some components in $\hat{\mathbf{e}}$.
where $\odot$ represents the Hadamard product. Compared to Equation (\ref{eq:problem}), we introduce additional indicator parameters $\boldsymbol{\alpha}$ which have the same size as $\hat{\mathbf{V}}$. However, we do not attempt to directly optimize Equation (\ref{eq:pruned_loss}) but leverage $\boldsymbol{\alpha}$ to determine the importance of each embedding dimension. To be specific, we can measure the effect of the $j$-th dimension of the $i$-th field on the loss function independently by masking it while keeping embedding values of other dimensions unchanged, and measure the change of the loss value, which is formulated as below:

\begin{equation}
    \Delta \mathcal{L}_{i,j} = \mathcal{L}(\hat{\mathbf{V}}\odot  \mathbf{1}, \hat{\Theta}; \mathcal{D}) - \mathcal{L}(\hat{\mathbf{V}}\odot (\mathbf{1} - \boldsymbol{\epsilon}_{i,j}), \hat{\Theta}; \mathcal{D})
\end{equation}

where $\mathbf{1}$ is an all-1 matrix with dimension $\sum_{i}^{m}n_{i} \times d$, and indicator matrix $\boldsymbol{\epsilon}_{i,j} \in \{0, 1\}^{\sum_{i}^{m}n_{i} \times d}$ is a binary matrix with value zero everywhere except for the position on the $j$-th dimension of the $i$-th field. 
However, computing all $\Delta \mathcal{L}_{i,j}$ is prohibitively expensive, which requires $md$ forward passes over the dataset.
On the other hand, $\mathcal{L}$ is not differentiable with respect to $\boldsymbol{\alpha}$.
Inspired by the approximation in \cite{koh2017understanding,lee2018snip}, we relax the binary constraint of $\boldsymbol{\alpha}$ to continuous range $[0, 1]$ such that $\Delta \mathcal{L}_{i,j}$ can be approximated by the gradients $\partial \mathcal{L} / \partial \boldsymbol{\alpha}_{i,j}$ (denoted as $g_{i,j}(\hat{\mathbf{V}}, \hat{\Theta}; \mathcal{D}_{b})$) at the stationary point $\boldsymbol{\alpha}=\mathbf{1}$:
\begin{equation}
    \begin{split}
    \Delta \mathcal{L}_{i,j} &\approx g_{i,j}(\hat{\mathbf{V}}, \hat{\Theta}; \mathcal{D}_{b}) = \frac{\partial\mathcal{L}(\hat{\mathbf{V}}\odot \boldsymbol{\alpha}, \hat{\Theta}; \mathcal{D}_{b})}{\partial\boldsymbol{\alpha}_{i,j}}\bigg |_{\boldsymbol{\alpha}=\mathbf{1}} \\
    &=\lim_{\delta \to 0}\frac{\mathcal{L}(\hat{\mathbf{V}}\odot \boldsymbol{\alpha}, \hat{\Theta}; \mathcal{D}_{b}) - \mathcal{L}(\hat{\mathbf{V}}\odot (\boldsymbol{\alpha}-\delta \boldsymbol{\epsilon}_{i,j}), \hat{\Theta}; \mathcal{D}_{b})}{\delta}\bigg |_{\boldsymbol{\alpha}=\mathbf{1}} \label{eq:grad_g}
    \end{split}
\end{equation}

In this way, we could efficiently compute all $g_{i,j}(\hat{\mathbf{V}}, \hat{\Theta}; \mathcal{D}_{b})$ via only one forward-backward pass using automatic differentiation on a mini-batch of dataset $\mathcal{D}$, denoted as $\mathcal{D}_b$. 
Here, a larger magnitude of $|g_{i,j}(\hat{\mathbf{V}}, \hat{\Theta}; \mathcal{D}_{b})|$ means that the corresponding dimension has a greater impact on the loss function (either positive or negative), and should therefore be retained.
Based on this hypothesis, the saliency score $\mathbf{s}_{i,j}$ of the $j$-th dimension of the $i$-th field is defined as the normalized magnitude of $|g_{i,j}(\hat{\mathbf{V}}, \hat{\Theta}; \mathcal{D}_{b})|$ as below:
\begin{equation}
    \mathbf{s}_{i,j} = \frac{|g_{i,j}(\hat{\mathbf{V}}, \hat{\Theta}; \mathcal{D}_{b})|}{\sum_{i=0}^m \sum_{j=0}^d |g_{i,j}(\hat{\mathbf{V}}, \hat{\Theta}; \mathcal{D}_{b})|}
    \label{eq:salience}
\end{equation}

After obtaining all saliency scores $s_{i,j}$, we rank them in descending order, and then sequentially remove dimensions with low saliency scores until the parameter budget $\kappa$ is reached. Precisely, the $\boldsymbol{\alpha}_{i,j}$ is computed as follows:
% the quantile value $\tilde{s}$ for parameter budget $\kappa$ is the $\kappa$-th top value of the sorted salience scores. Then 
\begin{equation}
    \begin{split}
    \boldsymbol{\alpha}_{i,j} &= \mathbb{I}(\mathbf{s}_{i,j} - \tilde{s} \geq 0), \forall i \in \{1,...,m\},j \in \{1,...,d\} \\
    &s.t.\ \Vert\boldsymbol{\alpha} \Vert_0 < \kappa ||\mathbf{V}||_{0}
    \end{split}
    \label{eq:prune}
\end{equation}

where $\mathbb{I(\cdot)}$ is the indicator function, and the quantile value $\tilde{s}$ is automatically determined based on the parameter budget.

It is worth noting that the above pruning method differs from the original network pruning method \cite{lee2018snip} in two ways. (1) We perform the single-shot pruning operation after pretraining, instead of prior to training as in the original one. The reason is that the embedding tables are independent of the model architecture, which prevents us from leveraging the architecture characteristics to perform pruning at the initialization stage. (2) Unlike the original one performing pruning at the weight level, which is analogous to pruning at the feature level in our context, we prune the embedding at the field level. The reason is that pruning at the feature level requires that the limited data in $\mathcal{D}_{b}$ must cover all unique features, which is hard to be satisfied due to sparsity and long-tail characteristics of features in recommender systems.

\subsubsection{\textbf{Retraining}}
After pruning, the mixed-dimensional embedding table $\bar{\mathbf{V}}=\{\bar{\mathbf{V}}_{1},\cdots,\bar{\mathbf{V}}_{q}\}_{q \le m}$ \footnote{$q \le m$ means that all dimensions of some fields are pruned, which also demonstrates that our method could be utilized to select important features automatically.},  $\bar{\mathbf{V}}_{i} \in \mathbb{R}^{d_{i} \times n_{i}}$ , are automatically obtained, where $d_{i}$ is the searched dimension for the $i$-th field, and $q$ is the number of fields after pruning. 
% we have obtained the sparse embeddings $\hat{\mathbf{e}}$. However, there exist many zero elements in $\hat{\mathbf{e}}$, which can be discarded to save the memory usage. We call this process as "resize", after which the varied-size embeddings $\tilde{\mathbf{e}}$ with non-zero elements alone are generated. 
However, traditional recommendation models such as FM \cite{rendle2010factorization} and DeepFM \cite{guo2017deepfm} require the dimensional consistency among input embeddings due to feature interaction operations (e.g., the dot product). In order to make the pruned embeddings be seamlessly integrated into various model architectures. We need to align embedding dimensions among different fields, which has been studied in previous works. For example, FmFM \cite{sun2021fm2} introduces additional $\frac{q(q-1)}{2}$ matrices to align the dimensions for each pair of embeddings from different fields. Differently, inspired by the work in \cite{cheng2020differentiable, zhao2020memory}, we utilize a simple yet effect method to align the dimension. Specifically, we only introduce $q$ field-wise transform matrices $\mathbf{M} = \{\mathbf{M}_1,...,\mathbf{M}_q\},\mathbf{M}_i \in \mathbb{R}^{d_{max} \times d_{i}}$, where $d_{max}=max(d_{1},\cdots,d_{q} )$ is the aligned embedding dimension equaling to the maximum dimension among all searched dimensions. Thus, we can get the aligned embedding $\bar{\mathbf{e}}_i \in \mathbb{R}^{d_{max}}$ for feature $\mathbf{x}_{i}$ as below:
\begin{equation}
    \bar{\mathbf{e}}_i = \mathbf{M}_{i} \bar{\mathbf{V}}_{i} \mathbf{x}_{i} \label{eq:align}
\end{equation}

Notice that we only introduce a tiny number of additional parameters, far less than those reduced embeddings. What's more, $\mathbf{M}_i$ can boost the representation of the pruned embeddings $\bar{\mathbf{V}}$: 1) they are projected into a high-dimensional space with more perspectives for expression 2) $\mathbf{M}_i$ is shared for all feature instances in field $i$, and thus their common attributes can be further modeled. After alignment, the dot product operation $<\cdot>$ of traditional recommendation models between features in the $i$-th field and the $j$-th field can be performed as follows:
% will be altered from $\mathbf{v}_{i,j}$ in Equation (\ref{eq:dfm_inc}) to $\tilde{\mathbf{v}}_{i,j}$, which is computed as:
\begin{equation}
    <\bar{\mathbf{e}}_i, \bar{\mathbf{e}}_j> \label{eq:aligned_inc}
\end{equation}

Besides dot product operation, the pruned embeddings are also easily adaptable to other feature interaction operations. For example, the DNN layer \cite{guo2017deepfm,cheng2016wide} only needs to adjust the dimension of the input layer to match the dimension of the output of the embedding layer. 
% Some other feature interaction layers will be slightly affected by the resizing process. To be specific, for MLP layers, the inputs are altered from $\mathbf{e}$ to $\tilde{\mathbf{e}}$, thus some connections between the first MLP layer and the zero elements in $\mathbf{e}$ are cut off.

After aligning, we need to retrain the pruned embedding table $\bar{\mathbf{V}}$, transform matrices $\mathbf{M}$, and other model parameters $\bar{\Theta}$ of the resulting slim model. According to the Lottery Ticket Hypothesis \cite{frankle2018lottery}, there exists a sub-model whose accuracy can match the original model after training for the same number of steps from scratch, and such a sub-model is called the \textit{Winning Ticket}. Moreover, compared to random initialization, restoring the parameters of the sub-model from the original well-trained model can further boost the performance. This theory inspires us to initialize the slim model using the pruned embedding table $\hat{\mathbf{V}}$, and randomly initialize other model parameters, i.e., $\mathbf{M}$. 

\begin{algorithm}
\caption{The proposed SSEDS algorithm}\label{alg:SSEDS}
\begin{algorithmic}
\Require Training dataset $\mathcal{D}$, base model $f(V,\Theta)$, parameter budget $\kappa$, and loss function $\mathcal{L}$. 
% \Comment{Equation (\ref{eq:pruned_loss})}
\Ensure Mixed-dimensional embedding table $\mathbf{V}^{*}$ s.t. $\Vert \mathbf{V}^{*} \Vert_0 < \kappa$ $\Vert \mathbf{V} \Vert_0$

\State \textbf{Pretraining:}
\State Optimize $\mathbf{V}$ and $\Theta$: $\hat{\mathbf{V}}$, $\hat{\Theta}$ $\xleftarrow{f}$ $\mathbf{V}$, $\Theta$ by the base model $f$.

\State \textbf{Single-shot Pruning}
\State Sample a mini-batch dataset $\mathcal{D}_{b} \sim \mathcal{D}$
\State Identify the saliency score $\boldsymbol{s}_{ij}$ for the $j$-th dimension of the $i$-th field \Comment{ Equation (\ref{eq:salience}) }
\State Rank $\boldsymbol{s}_{ij}$ in descending order
\State Prune embedding table $\bar{\mathbf{V}} \xleftarrow{} \hat{\mathbf{V}}$ \Comment{ Equation (\ref{eq:prune}) }

\State \textbf{Retraining}
\State Align the embedding dimension  \Comment{Equation (\ref{eq:align})}
\State Retrain the slim model ${\mathbf{V}^{*},\mathbf{\Theta}^{*}} \xleftarrow{} \bar{\mathbf{V}},\bar{\Theta}$

% \While{$N \neq 0$}
% \If{$N$ is even}
%     \State $X \gets X \times X$
%     \State $N \gets \frac{N}{2}$  \Comment{This is a comment}
% \ElsIf{$N$ is odd}
%     \State $y \gets y \times X$
%     \State $N \gets N - 1$
% \EndIf
% \EndWhile
\end{algorithmic}
\end{algorithm}

\section{Experiments}
In this section, we conduct extensive experiments aiming to answer the following research questions (RQs):
% To validate the effectiveness of the proposed method, we conduct offline experiments on two public datasets as well as an online A/B test on the WeChat Subscription platform. All these experiments are about the CTR prediction task, which aim to answer the following research questions (RQ):
\begin{itemize}
    \item \textbf{RQ1:} How does the proposed SSEDS perform under the different parameter budgets compared to other state-of-the-art methods?
    \item \textbf{RQ2:} Can the proposed SSEDS improve both the search efficiency of embedding dimensions and the inference efficiency of the model?
    \item \textbf{RQ3:} How does the proposed SSEDS perform on the practical recommender system in the industry?
    \item \textbf{RQ4:} How do the different components (e.g., retraining and \textit{Winning Ticket}) affect the performance of the proposed SSEDS?
    \item \textbf{RQ5:} Is it necessary to prune most of the embeddings?
\end{itemize} 

\subsection{Experimental Setups}

\subsubsection{Datasets}
\begin{table}[htbp]
  \centering
  \caption{The statistical information of datasets.}
    \begin{tabular}{cccc}
    \toprule
    Dataset & \#Instances & \#Fields & \#Features \\
    \midrule
    \midrule
    Criteo & 46M & 39    & 1M \\
    Avazu & 40M & 22    & 0.6M \\
    % WeChat Subcription &   1.1M  & 62 & 2.3M  \\
    \bottomrule
    \end{tabular}%
  \label{tab:addlabel}%
\end{table}%

The offline experiments are conducted on two widely used public datasets (i.e., Criteo and Avazu). Both of them are randomly split into training/validation/test sets as ratio 8:1:1,
and the detailed information of datasets is summarized in Table 1.
\begin{itemize}
    \item \textbf{Criteo\footnote{https://www.kaggle.com/c/criteo-display-ad-challenge}:} It is a real-world industry dataset for CTR prediction, which consists of 45 million users' click records on ads over one month. Each click record contains 13 numerical feature fields and 26 categorical feature fields. Following the winner of Criteo Competition\footnote{https://www.csie.ntu.edu.tw/20r01922136/kaggle-2014-criteo.pdf}, we group those low-frequency (less than 10) features as a single feature "others", and the numerical feature is transformed by $log^{2}(x),\ if\  x>2$.    
    \item \textbf{Avazu\footnote{https://www.kaggle.com/c/avazu-ctr-prediction}:} It consists of 40 million users' click records on ads over 11 days, and each record contains 22 categorical features. We use the pre-processing method as Criteo for the low-frequency features (less than 10) in Avazu.
    % \item \textbf{WeChat Subscription:} It consists of 1.1 million users' click records on WeChat Subscription videos sampled over 7 days, and each record contains 62 categorical features.
\end{itemize}

\subsubsection{Baselines}
We compare the proposed SSEDS with the following state-of-the-art methods including uniform-dimensional methods such as FM \cite{rendle2010factorization}, DeepFM \cite{guo2017deepfm} and Wide$\&$deep \cite{cheng2016wide}, as well as different kinds of EDS methods including the heuristic based method (i.e., MDE \cite{ginart2021mixed}), the HPO based method (i.e., AutoDim), and embedding pruning based methods (i.e., PEP \cite{liu2021learnable} and Deeplight).
\begin{itemize}
    \item \textbf{Base methods:} We select representative CTR models including FM \cite{rendle2010factorization}, DeepFM \cite{guo2017deepfm} and Wide$\&$deep \cite{cheng2016wide} as base models. These methods utilize uniform embedding dimension for all feature fields. 
    \item \textbf{MDE \cite{ginart2021mixed}:} MDE (short for Mixed Dimension Embedding) is the method that heuristically sets embedding dimensions based on features' popularity by human-designed rules.
    \item \textbf{AutoDim \cite{zhao2020memory}} AutoDim utilizes a soft and continuous manner to calculate the weights over various dimensions for feature fields. Then the embedding architecture is derived from the maximum weight.
    \item \textbf{PEP \cite{liu2021learnable}:} PEP is an embedding pruning based method for embedding dimensions search, which utilizes learnable thresholds to automatically prune redundant feature parameters so that sparse embeddings can be obtained adaptively.
    \item \textbf{Deeplight \cite{deng2021deeplight}} Deeplight proposes to prune both parameters in the embedding and DNN layers to solve the high-latency issues in CTR prediction.  
\end{itemize}

\begin{figure*}[htbp]
\centering
\setlength{\abovecaptionskip}{0.2cm}
\includegraphics[width=1\textwidth]{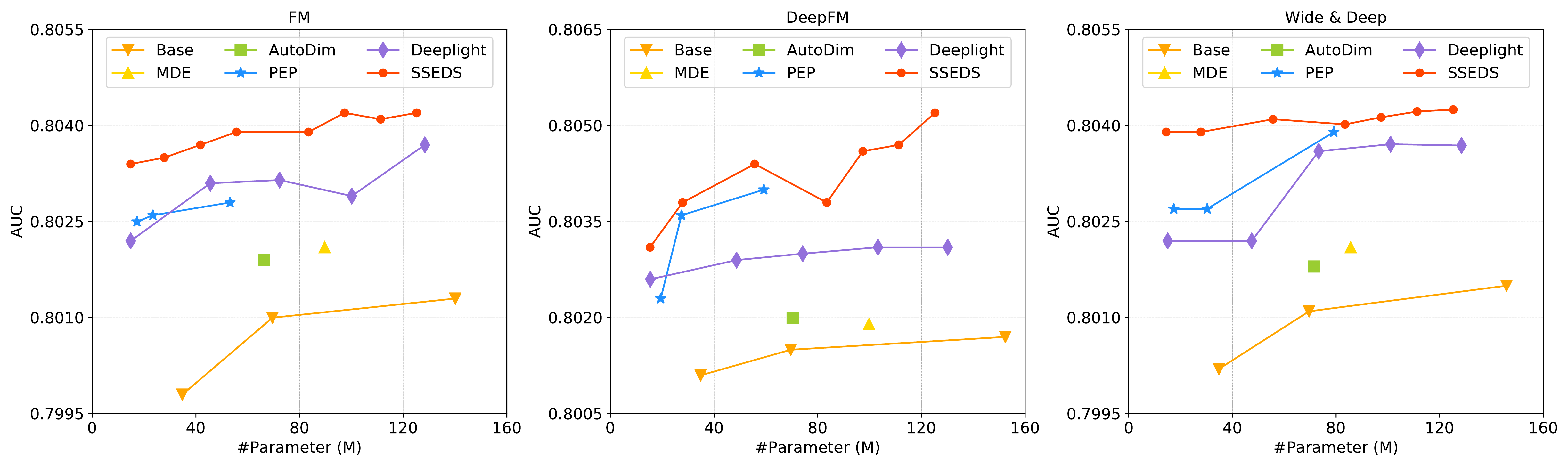}
\caption{The AUC-\#Parameter performance on the Criteo. (M=Million) } %最终文档中希望显示的图片标题
% \label{overview} %用于文内引用的标签
% \vspace{-0.3cm}
\end{figure*}

\begin{figure*}[htbp]
\centering
\setlength{\abovecaptionskip}{0.2cm}
\includegraphics[width=1\textwidth]{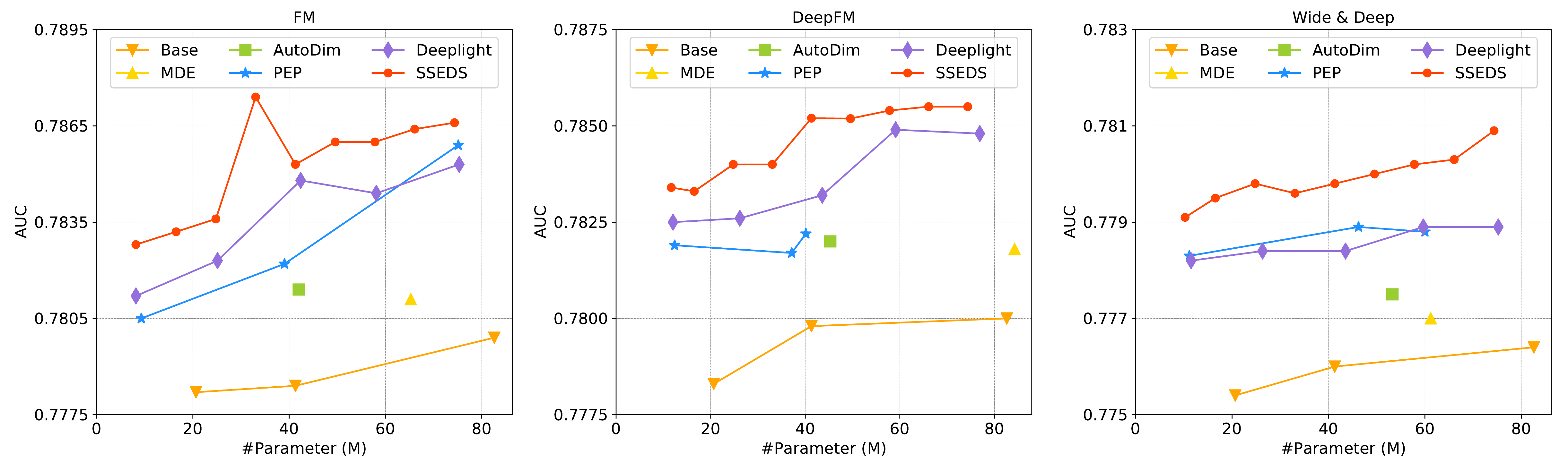}
\caption{The AUC-\#Parameter performance on the Avazu. (M=Million) } %最终文档中希望显示的图片标题
% \label{overview} %用于文内引用的标签
% \vspace{-0.3cm}
\end{figure*}

\subsubsection{Hyperparameters:} The hyperparameters of the proposed SSEDS are set as follows. The initial embedding dimensions (i.e., in the pretraining stage) are 128 for all feature fields, and the parameter budget $\kappa$ is 0.1 in the pruning stage. For deep layers, they contain 2 fully connected hidden layers with 1024 units in each layer, and the activation function for the layer's outputs is $ReLU(\cdot)$. We utilize Adam optimizer \cite{kingma2014adam} with an initial learning rate 0.001 throughout the experiments, and the batch size is set to 2048 for all datasets. The baselines (i.e., MDE, PEP, and Deeplight) are implemented by the codes provided by the authors. For a fair comparison, we set the initial (maximum) embedding dimension as 128 for all baseline models.
All the offline experiments are run on a single machine with 2 Tesla P100 GPU with 32G memory.

\subsubsection{Evaluation metrics:} The CTR prediction task can be viewed as the binary classification (i.e., click and not click) task. Thus, we utilize the common metric \cite{zhu2021towards}, i.e., AUC (Area Under the ROC Curve) to evaluate the recommendation performance of all methods. On the other hand, we also count the number of model parameters, denoted as $\#Parameter$, to measure model space complexities.

\subsection{CTR Prediction (RQ1)}
To validate the effectiveness of the proposed SSEDS, we compare it with other baselines on CTR prediction tasks. Specifically, the automated embedding dimension search methods (i.e., MDE, AutoDim, PEP, Deeplight, and proposed SSEDS) are regarded as plugins integrated into three base models (i.e., FM, DeepFM, and Wide$\&$Deep) respectively. Furthermore, in order to validate the model performance under different parameter budgets $\kappa$, we vary $\kappa$ from 0.1 to 0.9 with step size 0.1 and 0.2 for the proposed SSEDS and the Deeplight\footnote{For a fair comparison, we only use Deeplight to prune embedding tables, while the parameters in the DNN component and the field matrix are not pruned.} respectively. For the PEP, we report its performance under three different sparsity levels. We only report the final optimized results for MDE and AutoDim due to their model characteristics. For base models, we set dimensions as $\{32,64,128\}$, and report their performance. The experimental results are shown in Figure 2 and Figure 3 for Criteo and Avazu, respectively, from which we can observe that:
\begin{itemize}
    \item As the dimension increases, all base models can significantly achieve better performance on both datasets. It indicates that uniformly increasing dimensions for all feature fields could enhance the expression of the embeddings, resulting in better recommendation performance.
    \item The EDS based methods (i.e., MDE, AutoDim, PEP, Deeplight, and SSEDS) perform better than uniform embedding dimension based methods (i.e., FM, DeepFM, and Wide\&Deep), which identifies that assigning the same embedding dimension for all feature fields is sub-optimal.
    \item The embedding pruning based EDS methods (i.e., PEP, Deeplight, and SSEDS) can achieve better performance than the heuristic based method (i.e., MDE) and the HPO based method (i.e., AutoDim) in most cases on both datasets. The possible reason is that embedding pruning based methods measure the importance of different dimensions at a finer-grained level (i.e., the embedding level), rather than at an upper level (i.e., the dimension level) as in heuristic and HPO methods.
    \item SSEDS is capable of significantly improving the performance of all base models by integrating searched mix-dimensional embeddings into them. For example, it improves the performance of DeepFM by 1.7\textperthousand \ and 4.3\textperthousand with respect to AUC scores, while reducing 90\% embedding parameters on the Criteo and Avazu, respectively. It is worth noting that merely 1\textperthousand \ improvement is meaningful because of the large user base of businesses \cite{cheng2016wide, guo2017deepfm}. 
    \item Under the various parameter budgets, the proposed SSEDS outperforms other EDS based methods on both datasets. We attribute such advances to the proposed saliency criterion that could explicitly identify the importance of each embedding dimension. Thus, the overall performance demonstrates the effectiveness and superiority of the proposed SSEDS.
    % \item The SSEDS could achieve the high recommendation performance with very a small parameter budget (i.e., $\kappa=0.1$) on both datasets, which gives a further verification on the over-parameterized problem in uniform embedding dimension based recommendation methods.
    % \item For Criteo dataset, with the increasing of parameter budget $\kappa$, the mean AUC has a downward trend, but the variance gradually increases. 
    % \item Unexpectedly, for Avazu dataset, it does not obtain similar results as Criteo, which demonstrates that simply setting small parameter budget $\kappa$ might not obtain the optimal performance. 

\end{itemize}

\subsection{Efficiency Analysis (RQ2)}
\begin{figure}[htbp]
\centering
\setlength{\abovecaptionskip}{0.2cm}
\includegraphics[width=0.5\textwidth]{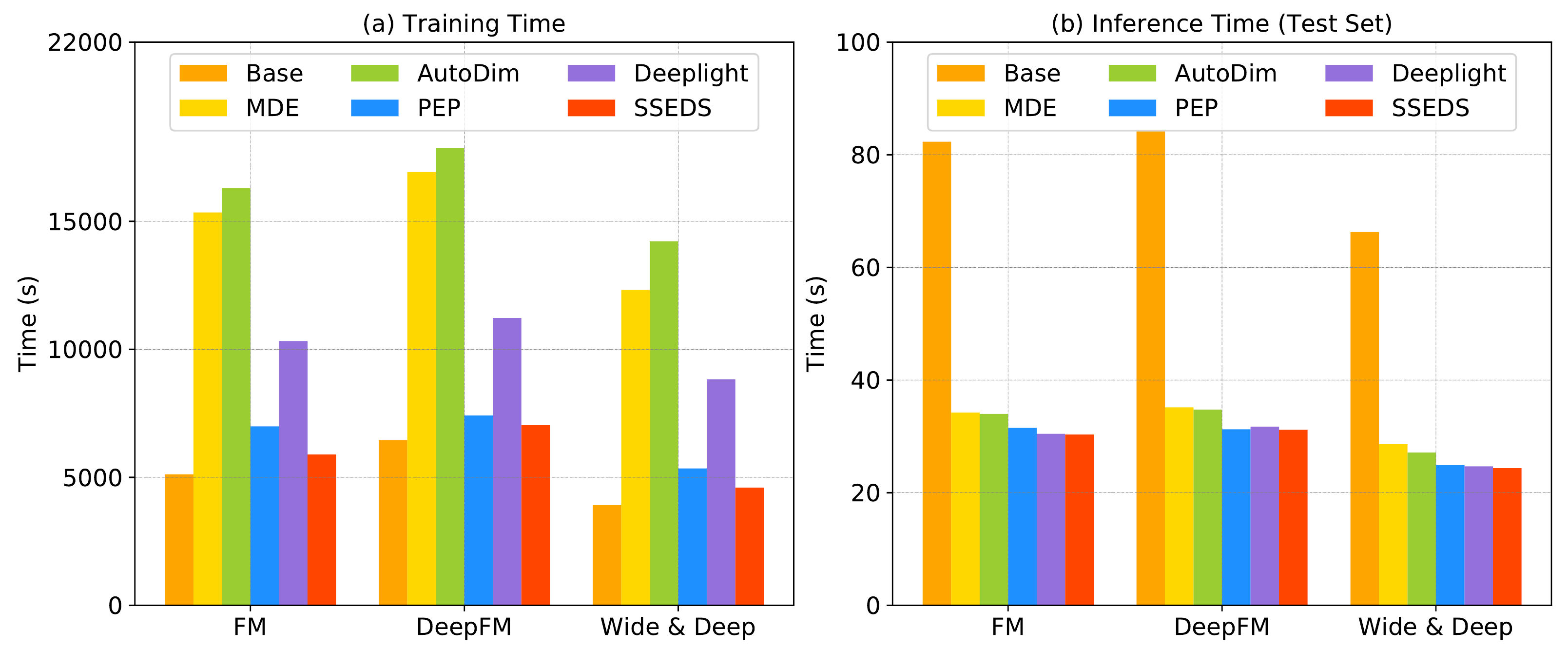}
\caption{The efficiency analysis of proposed SSEDS on Criteo. } %最终文档中希望显示的图片标题
\label{effcri} %用于文内引用的标签
% \vspace{-0.3cm}
\end{figure}

\begin{figure}[htbp]
\centering
\setlength{\abovecaptionskip}{0.2cm}
\includegraphics[width=0.5\textwidth]{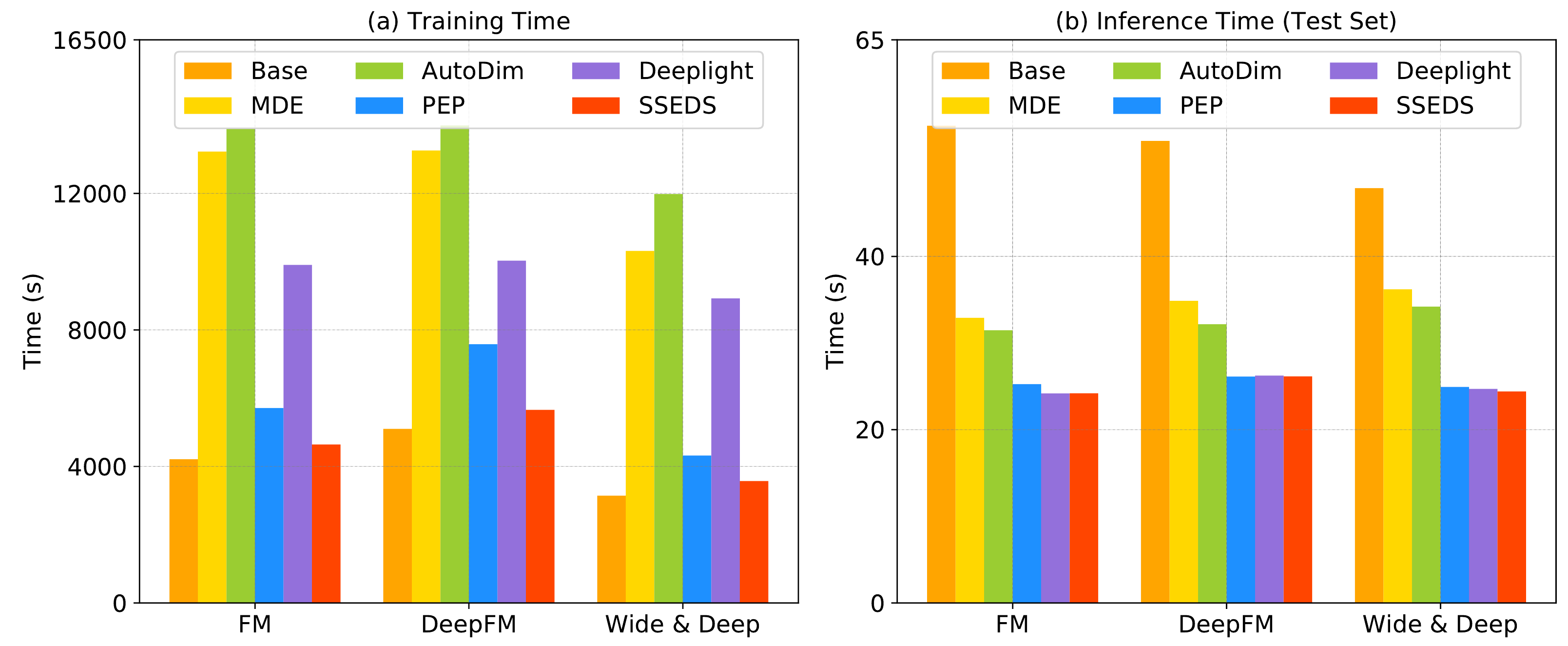}
\caption{The efficiency analysis of proposed SSEDS on Avazu. } %最终文档中希望显示的图片标题
\label{effava} %用于文内引用的标签
% \vspace{-0.3cm}
\end{figure}

It is inevitable to cost additional time to search for suitable embedding dimensions for different features. Thus, we study the time costs of the training stage (i.e., including pretraining, embedding dimension search, and retraining stages) on the training set, as well as the inference stage on the whole test set. Specifically, we set the training epochs as 3 for all methods for a fair comparison. Experimental results are shown in Figure \ref{effcri} and Figure \ref{effava} respectively, and we can find out that:
\begin{itemize}
    \item The proposed SSEDS spends the least additional time for the training stage, which is reasonable since SSEDS only need to prune the embeddings once after the pre-training, rather than to require iterative optimization procedures in other EDS methods.
    \item The EDS based methods could significantly reduce the inference time compared to base models due to reducing a large number of redundant embeddings. Furthermore, the inference time of SSEDS is comparable with other EDS methods due to the similar number of parameters. Thus, the overall performance of SSEDS in terms of efficiency, especially the training efficiency, is superior to other methods.
\end{itemize}

\subsection{Online A/B Test Experiment (RQ3)}
\begin{figure}[htbp]
\centering
\setlength{\abovecaptionskip}{0.2cm}
\includegraphics[width=0.5\textwidth]{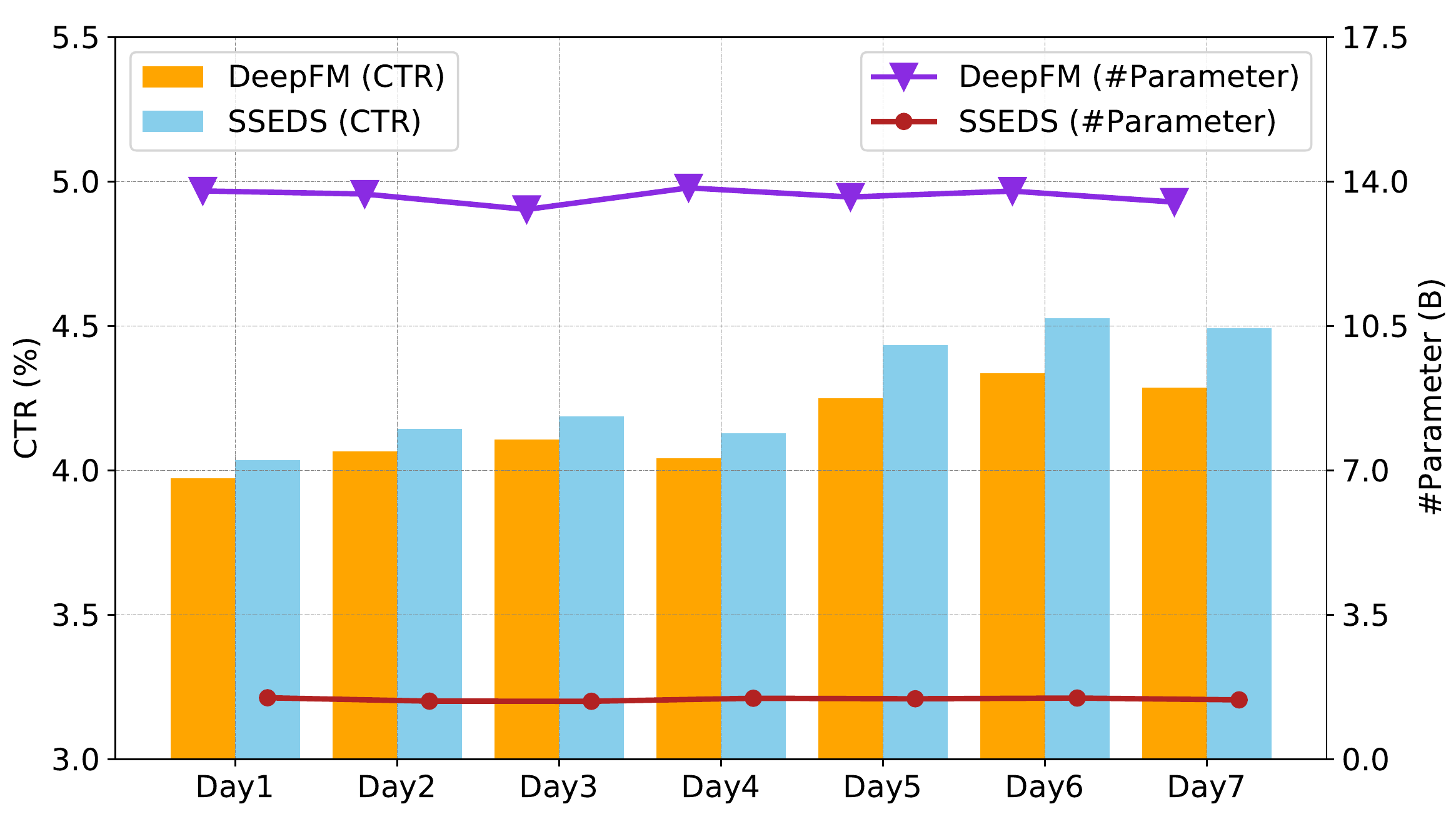}
\caption{The results of a 7-day online A/B test on WeChat Subscription platform. (B=Billion) } %最终文档中希望显示的图片标题
\label{online} %用于文内引用的标签
% \vspace{-0.3cm}
\end{figure}

To validate whether SSEDS could improve the performance while reducing the number of parameters on the practical recommender system in the industry, we deploy the SSEDS with a parameter budget $0.1$ on the WeChat Subscription recommendation platform that covers hundreds of millions of users and generates hundreds of billions of click records every day. This online recommendation service aims to recommend a set of videos that users are most likely to click. We conduct a 7-day online A/B test experiment on this platform. Specifically, the base model is DeepFM \cite{guo2017deepfm} that has already been deployed on the WeChat Subscription platform, where about 3\% of all users are set for the experimental group (SSEDS) and reference group (DeepFM), respectively. Both DeepFM and SSEDS are in an incrementally training manner every hour. We report the CTR (\%) values and corresponding the number of parameters of the two models for each day in Figure \ref{online}. From the results, we can observe that:
\begin{itemize}
    \item SSEDS could significantly improve about $4\%$ of online CTR metric while reducing nearly 90\% parameters, which confirms the practicality of SSEDS for the recommender system in the industry. 
    \item With the continually incremental training, there is a growing trend of improvement, which demonstrates that the proposed method could well capture the dynamic changes of the data distribution. It is meaningful because the data distribution constantly changes in a real industrial scenario.
\end{itemize}

\subsection{Ablation Study (RQ4)}
This subsection aims to explore the effects of retraining and the \textit{Winning Ticket}. The experiments are conducted on the Criteo and Avazu datasets for the CTR prediction task, and take DeepFM as the base model. In particular, we first explore the effect of retraining, denoted as \textit{SSEDS w/o retraining}, which implements the SSEDS without retraining. That is, we only use the embeddings from the pretrained model, and then prune the abundant dimensions via single-shot pruning. The randomly initialized transform matrices are utilized to align the dimension for different fields. We then explore the effect of the \textit{Winning Ticket}. Recall that we use the pruned embeddings (i.e., the \textit{Winning Ticket}) as the initial embeddings for the retraining stage. Thus, we explore its effect by random initialization in the retraining stage, denoted as \textit{SSEDS w/o ticket}. Other settings are the same as the above experiments. The experimental results are shown in Table \ref{tab:abla}, we can observe that:
\begin{itemize}
    \item SSEDS without retraining (i.e., SSEDS w/o retraining) degrades the model performance of SSEDS, but is still better than the base model on both two datasets. The possible reason is that over-parameterized embeddings might introduce noise or result in overfitting, which in turn demonstrates the necessity of embedding pruning.
    \item SSEDS retraining with random initialization (i.e., SSEDS w/o ticket) also degrades the model performance of SSEDS, but still achieves better performance than the base model and SSEDS w/o retraining. It implies that the \textit{Winning Ticket} could improve the recommendation performance of the model.
\end{itemize}
\begin{table}[htbp]
  \centering
  \caption{The ablation study results with respect to the retraining and the \textit{Winning Ticket}.}
    \begin{tabular}{ccc}
    \toprule
    Methods & Criteo & Avazu \\
    \midrule
    \midrule
    Base (DeepFM)  & 0.8017 & 0.78 \\
    SSEDS w/o retraining & 0.8020 & 0.7812 \\
    SSEDS w/o ticket & 0.8027 & 0.7826 \\
    \midrule
    SSEDS & \textbf{0.8031} & \textbf{0.7834} \\
    \bottomrule
    \end{tabular}%
  \label{tab:abla}%
\end{table}%

\subsection{Pruning Results Analysis (RQ5)}
\begin{figure}[htbp]
\centering
\setlength{\abovecaptionskip}{0.2cm}
\includegraphics[width=0.5\textwidth]{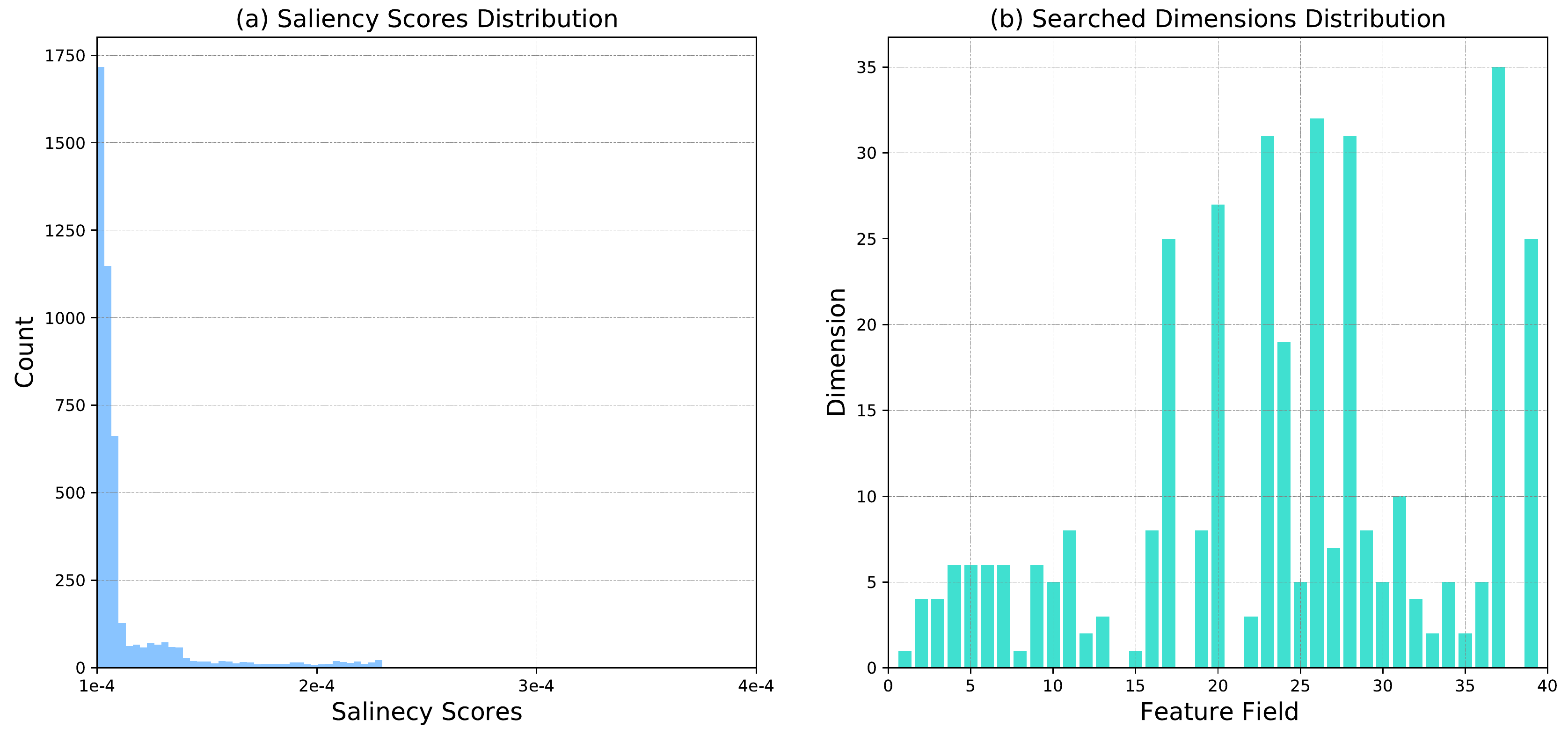}
\caption{The distribution of saliency scores and searched dimensions for the Criteo dataset. } %最终文档中希望显示的图片标题
\label{ssoncri} %用于文内引用的标签
% \vspace{-0.3cm}
\end{figure}

\begin{figure}[htbp]
\centering
\setlength{\abovecaptionskip}{0.2cm}
\includegraphics[width=0.5\textwidth]{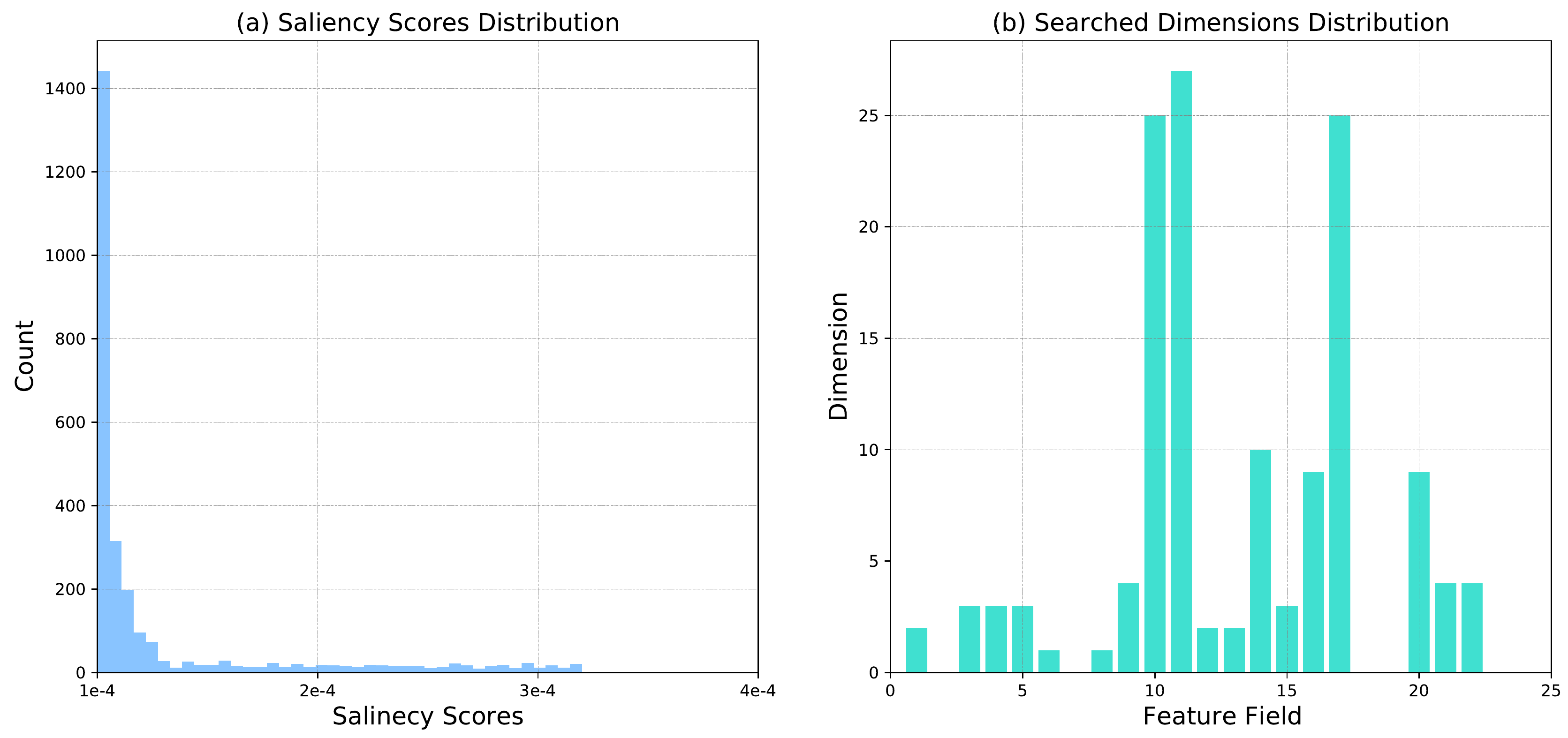}
\caption{The distribution of saliency scores and searched dimensions for the Avazu dataset. } %最终文档中希望显示的图片标题
\label{ssonava} %用于文内引用的标签
% \vspace{-0.3cm}
\end{figure}

To further demonstrate the necessity of embedding pruning, we present the distribution of saliency scores as well as the searched dimensions under the parameter budget $\kappa=0.1$. The experimental results are shown in Figure \ref{ssoncri} and Figure \ref{ssonava} for Criteo and Avazu datasets, respectively. We can observe that:
\begin{itemize}
    \item For both datasets, the saliency scores obey the power-law distribution, i.e., only a tiny proportion of dimensions have a large effect on the loss function. Such results are consistent with the characteristic of features in the recommender system, i.e., features following the long-tail distribution \cite{10.1145/3366424.3383416,qu2021imgagn}. Thus, it is reasonable and necessary to prune most of the embedding parameters.
    \item Only a small portion of feature fields are assigned with relatively large dimensions, while most of the fields are assigned small dimensions, or even 0 dimensions (meaning that the corresponding field is removed). It demonstrates that our method could also be utilized as an effective technique for automatic feature selection. 
\end{itemize}
% This subsection explores the effect of the hyperparameters. One crucial hyperparameter of SSEDS is the parameter budget $\kappa$, which determines the degree of sparsity of pruned embeddings. We conduct experiments on both Criteo and Avazu datasets for the CTR prediction task. In particular, we vary the size of $\kappa$ from 0.1 to 0.9 with step size 0.1, and for each budget size, we independently run SSEDS 5 times with randomly re-splitting datasets and random initialization. Other settings are the same as the above experiments. The performance of SSEDS with various parameter budgets $\kappa$ is shown in Figure 4. From the results, we can observe that:

\section{Conclusion}
In this paper, we have modeled the EDS problem in the recommender system as the embedding pruning problem, where the suitable embedding dimensions for different feature fields could be automatically obtained by removing those relatively unimportant dimensions under the desired parameter budget. To achieve this goal, we proposed a single-shot embedding dimension search method, termed SSEDS, which introduces a saliency criterion for identifying the importance of embedding dimensions in an efficient way. Extensive offline experiments have been conducted to validate the effectiveness and efficiency of the proposed SSEDS on two widely used datasets. The experimental results have shown that the proposed SSEDS can achieve better recommendation performance than both the traditional uniform dimension based methods and recent EDS based methods while reducing a large number of embedding parameters (about 90\%). On the other hand, the efficiency of embedding dimensions search for SSEDS is superior to other methods. Furthermore, the proposed SSEDS has also been deployed on the WeChat Subscription platform for online recommendation services. The 7-day A/B test results have demonstrated that SSEDS could significantly improve the performance of the online recommendation model while reducing resource consumption.

In the future, we plan to employ the proposed SSEDS to different recommendation tasks and explore how to leverage the prior knowledge of features to improve the model performance further. Furthermore, we will also explore to search embedding dimensions in a fine-grained manner, i.e., searching embedding dimensions for each unique feature, and dynamic situations \cite{qu2020continuous}. Finally, we will attempt to study the theoretical part of embedding pruning in order to guide the dimension assignment.

\section{Acknowledgement}
This work is partially supported by the Shenzhen Fundamental Research Program under the Grant No. JCYJ20200109141235597, National Science Foundation of China under grant No. 61761136008, Shenzhen Peacock Plan under Grant No. KQTD2016112514355531, Guangdong Introducing Innovative and Entrepreneurial Teams under grant No. 2017ZT07X386, Australian Research Council Future Fellowship (FT210100624) and  Discovery Project (DP190101985). The authors would like to thank Chong Liu for providing valuable suggestions on earlier version of this paper.

\bibliographystyle{ACM-Reference-Format}
\bibliography{main}

\end{document}